\documentclass{ws-procs975x65}

\begin{document}

%to switch ON running title
%\markboth{I. Bombaci}{Quark matter in compact stars}

%\wstoc{Quark matter in compact stars}{I. Bombaci}

\title{QUARK MATTER IN COMPACT STARS: 
ASTROPHYSICAL IMPLICATIONS AND POSSIBLE SIGNATURES}

\author{IGNAZIO BOMBACI}

\address{Dipartimento di Fisica "Enrico Fermi", Universit\`a di Pisa, and \\
INFN sezione di Pisa, Largo Bruno Pontecorvo, 3 
I-56127, Pisa, Italy\\
\email{bombaci@df.unipi.it}}

% WARNING. in standard latex cls file formatting, at this point
% \maketitle would typeset the above titlepage information
% but WS has chosen to be nonstandard and have each line typeset 
% as it is digested.
% no abstract is necessary.
% \bodymatter below resets the footnote counter and symbols after 
% possible use in the title matter.

\begin{abstract}
After a brief  non technical introduction of the basic properties of strange quark matter (SQM) 
in compact stars, we consider some of the late important advances in the field, and  discuss  
some recent astrophysical observational data that could shed new light on the possible presence 
of SQM in compact stars.  
We show that above a threshold value of the gravitational mass a neutron star (pure hadronic star)  
is metastable to the decay (conversion) to an  hybrid neutron star or to a strange star.  
We explore the consequences of the metastability of ``massive'' neutron stars  
and of the existence of stable compact ``quark'' stars (hybrid neutron stars or strange stars)  
on the concept of limiting  mass of compact stars, and we give an extension of this concept  
with respect to the {\it classical} one given in 1939 by Oppenheimer and Volkoff.        
\end{abstract} 

\bodymatter

\section{A brief history of quark matter in compact stars}\label{intro}
In 1964 Murray Gell-Mann\cite{gell-mann} and George Zweig\cite{zweig}  proposed that 
the proton, the neutron, and all the other hadrons ({\it i.e.} particles which feel 
the {\it strong interaction}) are not elementary, but are composed of smaller (point-like) constituents which were called {\it quarks}.
\footnote{Gell-Mann borrowed this name from a line of the novel  
{\it Finnegans Wake} by James Joyce.}   
In the original model there are three fundamental quarks nicknamed 
{\it up} ({\it u}) {\it down } ({\it d}) and {\it strange} ({\it s}).  
The premiere evidence for the existence of quarks came already at the end of 1967 
from the first high energy electron-proton scattering experiment\cite{slac} performed 
at the Stanford Linear Accelerator Center (SLAC),  and  a growing body of direct evidence 
for their  reality 
\footnote{Isolated quarks have never been observed. This has led to the quark confinement 
hypothesis, as one of the basic features of Quantum Chromodynamics (QCD), 
which is the fundamental theory of strong interactions.}    
was accumulated in the following years in numerous 
experiments in different laboratories all around the world.  

It was then natural to speculate that when nuclear matter is compressed to 
densities so high that nucleons substantially overlap, a new phase of matter, in which 
quarks are the relevant  degrees of freedom, could be formed. 
Already in 1965, Ivanenko e Kurdgelaidze\cite{ik65} suggested the possible existence of 
a ``quarkian core'' in very massive stars. 

Two years later, at the end of 1967,  Jocelyn  Bell discovered\cite{PSR}  the first pulsar. 
Pulsars  were soon interpreted\cite{pac-gold}  as strongly magnetized rotating 
neutron stars,  the compact stellar remnants of supernova explosions 
hypothesized\cite{bz} in 1934  by Baade and Zwicky \footnote{In 1937 
Landau suggested the existence of a ``neutron core'' (neutron matter core) in normal 
stars, to explain the energy source of stars as due to the release of gravitational energy 
via matter accretion  into the  stellar  neutron core.}    
soon after the Chadwick's discovery of the neutron.  
The first calculation of the structure on a neutron star was performed in 1939  by 
Oppenheimer and Volkoff\cite{ov39}.  They assumed the star to be composed 
by pure neutron matter described as a non-interacting relativistic Fermi gas. 
More sophisticated calculations, with the inclusion of the effects of 
nuclear interactions on the equation of state (EOS) of neutron matter, 
were available\cite{cam59} at the end of 1950s.  
A more refined model for the internal composition of neutron stars 
was introduced in 1960 by Ambartsumyan and Saakyan\cite{amb60}, 
which suggested the possible presence of hyperons 
($\Lambda$, $\Sigma^{-}$, $\Sigma^{0}$, $\Sigma^{+}$, $\Xi^{-}$ and $\Xi^{0}$ particles) 
in the inner core of neutron stars. 
All these early  models gave,  with a reasonable accuracy,  the gross properties
({\it i.e.}  maximum mass, radii, and central densities) of neutron stars. 
Particularly, it was clear that the central density of the maximum mass 
configuration could be as high as 10 times the central density 
($\rho_0 = 2.8 \times 10^{14} {\rm g/cm}^3$)  of heavy atomic nuclei.   
Thus it was quite evident at the end of the 1960s and in the early 1970s, 
that neutron stars were the best candidate in the universe where  
quark matter could be found.  
It was in those years that the idea of pure quark stars\cite{ito70} or hybrid hadronic-quark 
stars\cite{ik69}  (termed ``baryon-quarkian stars'' by the authors of ref. \cite{ik69}) was conceived.  
The earliest studies\cite{cp75,bc76,kk76}, however  indicated that it was unlikely that quark matter 
could be found in stable neutron stars, or to have a third family of quark compact stars.  
This conclusion was mainly due to some simplification in the treatment 
of the quark-deconfinement phase transition, which was considered as one at 
constant pressure. 
Further investigations\cite{gle92,gle_book} have established that,  since  
neutron star matter is a multicomponent system with  two conserved 
``charges'' ({\it i.e.,} electric charge and baryon number), 
the quark-deconfinement phase transition proceeds through a mixed phase over a 
finite range of pressures and densities according to the Gibbs' criterion for phase equilibrium.  
These \cite{gle92,gle_book} and subsequent studies have established that 
neutron stars may, very likely, contain quark matter in their interiors. 
Neutron  stars  which possess a quark matter core either as a mixed phase
\footnote{see however ref.s\cite{heisel93,voskr02} for later studies on the mixed hadron-quark phase, 
where the effects of Coulomb, surface energies, and charge screening are taken into account.}
of deconfined quarks and hadrons or as a pure quark matter phase are called 
{\it hybrid neutron stars} or shortly {\it hybrid stars}\cite{gle_book}  (HyS).
The  more {\it conventional} neutron stars  in which no fraction of quark matter 
is present, are currently referred to as  {\it pure hadronic stars} (HS).  

A further crucial step in the study of quark matter in astrophysical contest was made by 
Arnold R. Bodmer in 1971.  In his pioneering work\cite{bod71}, Bodmer suggested 
the possibility of {\it collapsed nuclei}. He conjectured that collapsed nuclei could 
have a lower energy than normal ({\it i.e.} made of protons and neutrons) nuclei, 
and he speculated that normal nuclei are very long-lived  {\it isomers} against collapse 
because of a ``saturation '' barrier between normal and collapsed nuclei.  
Among other possibilities, Bodmer discussed collapsed nuclei as composite 
states of 3A  {\it u}-{\it d}-{\it s} quarks (being A the baryon number of the nucleus). 
These many-body {\it u}-{\it d}-{\it s} quark states are what we call, with modern 
terminology,  {\it strangelets} or {\it quark nuggets}.
Bodmer also speculated that {\it collapsed nuclei may have been copiously produced 
in the initial extremely hot and dense stages of the universe},  and proposed that 
collapsed nuclei may condensate as peculiar very  compact massive ``black''  objects 
being possible candidates  for dark matter. 

The feasible stability of multi-quark systems was also investigated\cite{ter79}  by 
Hidezumi Terazawa, who termed them {\it super-Hypernuclei}. 
However, these ideas brought a wide attention by the scientific community 
only after Edward Witten published, in 1984,  his widely known paper\cite{witt84} on  the 
``cosmic separation of phases'', where he examined  the cosmological and  
astrophysical consequences of the {\it absolute stability} of strange quark matter 
(SQM){\footnote{see next section for the definition of strange quark matter.}, 
and calculated the bulk properties of {\it Strange Stars} within a simple model 
for SQM inspired to the MIT bag model for hadrons. 
In the following years the properties of SQM and the properties of strange stars 
were calculated by numerous groups\cite{afo86,hzs86,bh89}. 

 A recent valuable advance  in the comprehension of quark matter properties, 
that is having a substantial impact on the study of neutron star physics, 
is the  discovery of color superconductivity.   
Superconductivity, as it is well known, is a general feature of degenerate 
Fermi systems, which become unstable if there exist any attractive interaction 
at the Fermi surface.  As recognized by Bardeen, Cooper and Schrieffer (BCS) 
this instability leads to the formation of a condensate of Cooper pairs and 
to the appearance of superconductivity. 
In Quantum Chromodynamics (QCD) the quark-quark (qq) interaction is 
strongly attractive in many channels. 
This will lead to quark pairing and color superconductivity. 
This possibility was already  pointed out\cite{cp75,bar77,fra78}  
in the mid 1970s (see also ref.\cite{bl84}). 

Very recently the study of color superconductivity has become a research subject 
full of activity, since the realization that the typical superconducting 
energy gaps in quark matter may be of the order of $\Delta \sim 100$ MeV, 
thus much higher than those predicted in the early works. 

The phase diagram of QCD has been widely analyzed in the light of color 
superconductivity\cite{Raj-Wil00,Nard01,Alf01,Cas-Nard04,sho05,bub05,Schaf05,Rust-etal06,mann07}.    
Quarks, unlike electrons, have different flavors  ({\it u}, {\it d}, {\it s} for the quark matter  
relevant for neutron star physics) and have color as well as spin degrees of freedom, 
thus many different quark pairing schemes are expected.  
At asymptotic densities (in the asymptotic freedom regime where perturbative 
QCD is valid) the ground state of QCD is the so called color-flavor locked (CFL) phase, 
in which all three quark flavors participate to pairing symmetrically 
{\footnote{This phase of SQM is electrically charge neutral without any need for electrons.}.  
In the density regime relevant for neutron star physics (well below the 
asymptotic density regime) the ground state of quark matter is uncertain 
and many possible patterns for color superconductivity have been predicted.  
 
%%%%%%%%%%%%%%% Fig.1%%%%%%%%%%%%%
\begin{figure}[t]
\begin{center}
\psfig{file=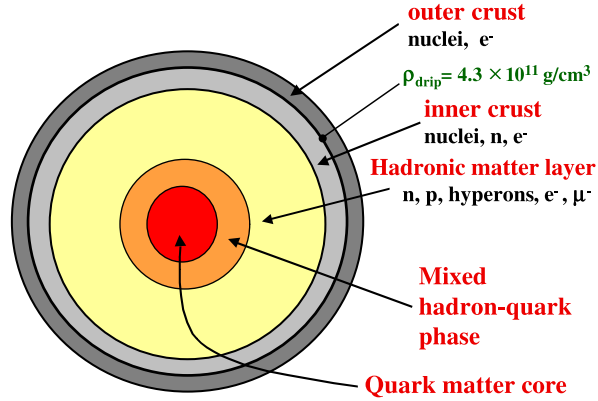,width=3.4in}
\end{center}
\caption{Schematic cross section of a hybrid star (see text for more details). } 
\label{aba:cross_sect_hybr}
\end{figure}
%%%%%%%%%%%%%%%%%%%%%%%%%%%%%%5

%%%%%%%%%%
\section{Hybrid stars}\label{hybr}
%%%%%%%%%%
A neutron star has a characteristic layered structure.  
The existence of various layers is a consequence of the onset of many different  
regimes  ({\it i.e.}  particle species and phases) of dense matter, 
which are expected in the stellar interior.  
The details of the stellar internal  structure  are sensitive 
to the particular high density EOS employed in the calculations. 
Very schematically, in the case of an hybrid star with sufficiently high mass (central density), 
these layers can be identified as follows:  

\noindent
(1) {\bf surface}  ($\rho_{Fe} = 7.9~{\rm g/cm}^3 < \rho < 10^6~{\rm g/cm}^3 $). 
Matter in the neutron star's surface is composed by a lattice of  
$^{56}$Fe nuclei (the endpoint of thermonuclear burning). 
As the density increases, more and more electrons detach from their 
respective nuclei, and at  $\rho \sim 10^4 {\rm g/cm}^3$, one has complete 
ionization. 

\noindent
(2) {\bf outer crust} ($10^6~{\rm g/cm}^3 < \rho < \rho_n^{drip}$). 
The outer crust is a solid layer consisting of a Coulomb lattice of heavy 
nuclei ($A>56$) in $\beta$--equilibrium with a degenerate relativistic 
electron gas. 
Going deeper into the crust (to higher density regions) nuclei in the lattice 
become more and more neutron rich, because the electron capture processes  
 $ e^-  + p  \rightarrow n + \nu_e$, lower the total energy of the system. 
 When the density reaches the value $\rho_n^{drip} \simeq  4.3 \times 10^{11}~{\rm g/cm}^3$ 
(neutron drip density), nuclei are so neutron rich that neutron states in 
the continuum begin to be filled. 

\noindent
(3) {\bf inner crust} ($\rho_n^{drip} < \rho < \rho_{core} $). 
Above the neutron drip density, the crust consist of a lattice of neutron rich 
nuclei embedded in a ultra--relativistic electron gas and in  a neutron gas. 
Due to the nuclear pairing force, neutrons forms Cooper pairs and are expected 
to be in a $^1S_0$ superfluid state.  
As the density of matter increases and approaches that of uniform nuclear matter
($\rho_{NM} \simeq  2.8 \times 10^{14}~{\rm g/cm}^3$),  one has a 
{\bf crust-core transition layer} consisting of a mixed phase (the so called 
 ``nuclear pasta'' phase)  of  ``exotic nuclei'' and a neutron--electron gas.  
First one  encounters a stratum composed of spherical blobs of nuclear matter 
(neutron rich super-heavy nuclei) embedded in the neutron and electron gas. 
With increasing density, the nuclear matter spherical blobs, first turn in 
rod-like structures (``spaghetti''),  and then to plate-like ones (``lasagna''). 
For higher densities one has the complementary nuclear shapes, {\it i.e.}, 
the so called ``anti-spaghetti'', and ``Swiss cheese'' (a phase in which balloons filled 
by neutron and electron gases are surrounded by nuclear matter).  
The appearance of nuclear pasta phase  in the inner crust, 
is due to finite size effects (particularly the surface and Coulomb energies) 
which settle the minimum of the the local total energy per baryon  of these  
competing exotic geometrical nuclear structures between each others and 
uniform $\beta$-stable nuclear matter.  

\noindent
(4) {\bf outer core} (pure hadronic layer)   
($\rho_{core} < \rho < \rho_{mix} $). 
When the density reaches  a value $\rho_{core} = $ 0.5 -- 0.8 $\rho_{NM}$,    
nuclei merge together and a phase transition to nuclear matter takes place 
In this region (nuclear matter layer) the star consist of asymmetric nuclear matter in 
$\beta$--equilibrium with $e^-$ and $\mu^-$.   
Neutron--neutron pairs form in a $^3P_0$ superfluid state and proton--proton 
pairs are in a $^1S_0$ superconducting state.   
Going deeper in the outer core, other hadronic constituents are expected, as 
hyperons,  or possibly  a Bose-Einstein condensate of negative kaons ($K^-$). 
 
\noindent
(5) {\bf mixed hadron--quark layer}  ($\rho_{mix} < \rho < \rho_{QM}$).   
At the critical density for the onset of the quark deconfinement phase transition, 
one encounters a layer consisting of a mixed phase between deconfined quarks and 
hadrons\cite{gle92,heisel93,gle_book}.   
On the top of this layer one has a Coulomb lattice of quark matter droplets embedded in a sea 
of hadrons and in a roughly uniform sea of electrons and muons.  
Moving toward the stellar center, various geometrical shapes (rods, plates) 
of the less abundant phase immersed in the dominant one are expected.  
This structured mixed phase ({\it quark pasta} phases) is analogous to {\it nuclear pasta} phase 
in the inner crust. 
It is important to stress that a number of recent studies\cite{heisel93,voskr02} have reexamined 
the physical conditions for the occurence of this structured mixed phase in a compact star. 
It has been shown\cite{heisel93,voskr02} that the formation of the mixed phase 
could be inhibited (or it can occur in a ``thin'' radial region of the star) for large values of the  
Coulomb and surface energies, or due to the effects of charge screening. 
 
\noindent
(6) {\bf quark matter core} ($\rho >  \rho_{QM} $).   
Finally, in the more massive stars, one could  find  a pure strange quark matter core. 
Different color superconducting phases are expected in this region of the 
star\cite{Raj-Wil00,Nard01,Alf01,Cas-Nard04,sho05,bub05,Schaf05,Rust-etal06,mann07}. 

\vskip 0.1cm                                                             
In Fig.~\ref{hybrid_comp}, we show the typical internal composition of a hybrid star.  
The cross section of the star has been obtained using a relativistic mean field EOS\cite{gm91}  
(model GM3) and is relative to the maximum mass configuration   ($M = M_{max} = 1.448 M_\odot$, 
with radius $R=10.2$~km, and central density  $\rho_c =  28. 1 \times 10^{14}$~g/cm$^3$).   
This star has a pure QM core which extend for about 2.5 km,  next it has a 
hadron-quark mixed phase layer with a thickness of about 5.5 km, 
followed by a nuclear matter layer  about 2 km thick.   
On the top we have the usual neutron star crust.       
The presence of quarks makes the EOS softer with respect to the corresponding  
pure hadronic matter EOS. The stellar sequence associated to the latter EOS  has the 
maximum mass configuration:  $M_{max} = 1.552~ M_\odot$, $R=10.7$~km,  
$\rho_c =  25.4 \times 10^{14}$~g/cm$^3$.   
%%%%%%%%%%%%%%% Fig.1%%%%%%%%%%%%%
\begin{figure}[t]
\begin{center}
\psfig{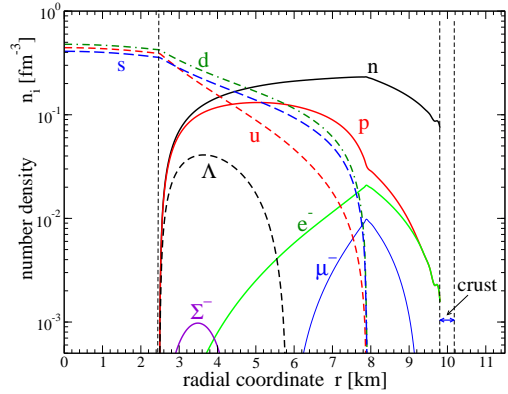}
\end{center}
\caption{The internal composition of an hybrid star with a mass 
$M$=$M_{max}$=$1.448~ M_\odot$.  The GM3 EOS \cite{gm91}  has been used for 
the hadronic phase,  and the bag model EOS,    
with $B = 136.6$~MeV/fm$^3$ and $m_s=150$ MeV, for the quark phase.}
\label{hybrid_comp}    
\end{figure}
%%%%%%%%%%%%%%%%%%%%%%%%%%%%%%

Many possible astrophysical signals for the presence of a quark core 
in neutron stars have been proposed\cite{GPW97,chub00}.  
Particularly,  pulse timing properties of pulsars have attracted much attention since 
they are a manifestation of the rotational properties of the associated neutron star. 
The onset of quark-deconfinement in the core of the star, will cause a change in the 
stellar moment of inertia\cite{GPW97}.  
This change will produce a peculiar evolution of the stellar rotational period  
($P= 2\pi/\Omega$) which will cause large deviations\cite{GPW97,chub00}  of the so 
called pulsar {\it braking index} $n(\Omega) = (\Omega \ddot{\Omega}/\dot{\Omega}^2)$ 
from the {\it canonical} value $n=3$, derived within the magnetic dipole  model for 
pulsars\cite{shap-teuk} and assuming a constant moment of inertia for the star.  
The  possible measurement of a value of the braking index very different 
from the canonical value ({\it i.e.} $ |n| >> 3$) has been proposed \cite{GPW97}   
as a signature for the occurrence of the quark-deconfinement phase transition 
in a neutron star.  
However, it must be stressed that a large value of the braking index 
could also results from the pulsar magnetic field decay or 
alignment of the magnetic axis with the rotation axis \cite{TK2001}.

%%%%%%%%%%
\section{Strange Quark Matter}\label{sqm} 
%%%%%%%%%%
Strange quark matter (SQM) is  an infinite deconfined mixture of {\it u}, {\it d} and {\it s} 
quarks  in a color singlet state (colorless),  together with an appropriate number 
of electrons to guarantee electrical neutrality.  
The compositions of SQM, for any given value of the total baryon number density $n$, 
is  determined by the requirement of charge neutrality and equilibrium with respect 
to the weak processes:  
\begin{equation}
                    u + e^-   \rightarrow d + \nu_e~,   ~~~~~~~~
                    u + e^-   \rightarrow s + \nu_e~, ~~~~~~~~
                    d   \rightarrow  u + e^- + {\bar\nu}_e ~,  \nonumber
\end{equation}
\begin{equation}     
                          s     \rightarrow u + e^- + {\bar\nu}_e ~,  ~~~~~~~~~~~~~~         
                          s +  u    \rightarrow d  + u                 \nonumber
\end{equation}
which can be expressed in terms of the various particles chemical potentials $\mu_i$ as  
\begin{equation}
 \mu_d = \mu_s  \equiv \mu   = \mu_u + \mu_e  
\label{chem_eq}
\end{equation}
in the case neutrino--free matter ($\mu_{\nu_e} =  \mu_{{\bar\nu}_e} = 0$). 

The charge neutrality condition requires:
\begin{equation}
{{2}\over{3}} n_u  - {{1}\over{3}} n_d - {{1}\over{3}} n_s - n_e = 0 \, ,
\label{charge_neut}
\end{equation}
where $n_i$  ($i= u, d, s, e$) is the number density for the different particle species.   
Equations ~(\ref{chem_eq}) and (\ref{charge_neut}) implies that there is only one 
independent chemical potential, which we call the {\it quark chemical potential} $\mu$. 
Due to the well known difficulties in solving the QCD equations on the lattice at 
finite density, various models (which incorporates the basic features of QCD) 
have been developed to get the equation of state of SQM.  
Along these lines, numerous studies (see {\it e.g.}  ref.\cite{bub05,blasch05,rust06,abu06,law06}
and references therein quoted) of SQM have been done using different variants 
of the Nambu--Jona-Lasinio model\cite{njl}. 
In some recent work\cite{gomez02,duhau04,gomez06,blaschke07} , 
nonstrange quark matter has been described using a more 
elaborate  approach based on a nonlocal covariant chiral quark model. 
With this EOS  various properties of hybrid stars have been calculated\cite{blaschke07}.

Here, we outline a schematic model\cite{bc76,fl78,fj84} for the equation of state of SQM, 
which is inspired to the MIT bag model for hadron. 
Despite this model is applicable at asymptotic densities (where perturbative QCD is valid), 
it has become very popular in the study of SQM in 
astrophysics\cite{gle_book,madsen,bomb01,weber}.     
The basic idea  of the model is to suppose that quarks are confined within a  
spherical region (the {\it bag}) of the QCD vacuum. Inside the bag, quarks interact 
very weakly (perturbatively) each other. The vacuum inside the bag (perturbative vacuum) 
is considered as an excited state of the true QCD vacuum outside the bag.    
Perturbative vacuum is characterized by a constant energy density $B$, the  {\it bag constant},  
which accounts in a phenomenological way of nonperturbative aspects of  QCD.  
This gives rise to an inward pressure $P_B=-B$ on the surface of the bag, 
which balances the outward pressure originating from the Fermi motion of quarks and 
from their perturbative  interactions.  
Thus in the MIT bag model for SQM, the essential phenomenological features of 
QCD, {\it i.e.} quark confinement and asymptotic freedom, are postulated from the beginning. 
 The short range qq interaction can be introduced in terms of a perturbative 
expansion in powers of the QCD  structure constant $\alpha_c$.  
The {\it up} and {\it down} quarks  are assumed to be massless ($m_u = m_d = 0$), 
and the {\it strange} quark to have a finite  mass, $m_s $, which  is taken  as 
a free parameter
\footnote{The value of the {\it current quark mass}, as reported by the Particle Data Group  
(http://pdg.lbl.gov/) are the following: $m_u = $ 1--3 MeV,    $m_d = $ 3--7 MeV, 
$m_s = 95 \pm 25$ MeV.}.         

The grand canonical potential $\Omega$  per unit volume of SQM, up to linear terms 
in $\alpha_c$ can be written as\cite{bc76,fl78,fj84,gle_book} 
\begin{equation} 
         \Omega = \Omega^{(0)} + \Omega^{(1)} + B   + \Omega_e \, .
\end{equation}
$\Omega^{(0)}$ is the contribution  of a non--interacting  {\it u,d,s} Fermi gas  
\begin{equation}
        \Omega^{(0)} = \Omega_u^{(0)} + \Omega_d^{(0)} + \Omega_s^{(0)} 
\end{equation}
\begin{equation}
        \Omega_q^{(0)} = - {1\over {(\hbar c)^3}} {1\over{4\pi^2}} {\mu_q}^4, 
                                              \qquad (q=u,d) 
\label{freegas_0}
\end{equation}
\begin{equation}
    \Omega_s^{(0)} = 
           - {1\over {(\hbar c)^3}} {1\over{4\pi^2}}
     \Bigg\{ \mu_s \mu_s^*  \Big(\mu_s^2 - {5\over 2} m_s^2\Big) + 
{3\over 2} m_s^4 \ln\Big({{\mu_s+\mu_s^*}\over{m_s}}\Big) \Bigg\} 
\end{equation}
with  $ \mu_s^*  \equiv  \Big(\mu_s^2 - m_s^2 \Big)^{1/2}$.    
The perturbative contribution, $\Omega^{(1)}$,  to the grand canonical potential density 
up to linear terms in $\alpha_c$  is 
\begin{equation}
  \Omega^{(1)} = \Omega_u^{(1)} + \Omega_d^{(1)} + \Omega_s^{(1)},  
\end{equation}
\begin{equation}
     \Omega_q^{(1)} = {1\over {(\hbar c)^3}} {1\over{4\pi^2}} 
       {{2 \alpha_c}\over {\pi}}  {\mu_q}^4,  \qquad (q=u,d) 
\end{equation}
\begin{eqnarray}
\Omega_s^{(1)}  &=& 
{1\over {(\hbar c)^3}} {1\over{4\pi^2}} {{2 \alpha_c}\over {\pi}} 
 \Bigg\{ 
3 \bigg[ \mu_s \mu_s^* - m_s^2 \ln\Big({{\mu_s+\mu_s^*}\over{m_s}}\Big)\bigg]^2 
  - 2 {\mu_s^*}^4         \nonumber \\
 &-&  3 m_s^4  \ln^2\Big({{m_s}\over{\mu_s}}\Big) 
    + 6  \ln\Big({{\rho_{ren}}\over{\mu_s}}\Big) 
\bigg[\mu_s \mu_s^* m_s^2  - m_s^4 \ln\Big({{\mu_s+\mu_s^*}\over{m_s}}\Big)\bigg] 
                                                     \Bigg\}  \nonumber \\  \,
\end{eqnarray}
where $\rho_{ren}$ is the so called {\it renormalization point} (see ref.\cite{fj84}). 
In the case of massless $u$ and $d$ quarks a standard choice is\cite{fj84}  
$\rho_{ren} = 313$~MeV.    
$\Omega_e$ is the electron contribution to $\Omega$.  Electrons are treated  as a massless 
(since in compact star interiors $\mu_e >> m_e$)  non-interacting Fermi gas, 
thus their contribution to $\Omega$ is given by Eq.~(\ref{freegas_0}) in terms of  $\mu_e$ .   
 
 To summarize, the EOS of SQM (at zero temperature) in the approximation   
for the grand potential density we are considering is: 
\begin{equation}
          P =  - \Omega^{(0)} - \Omega^{(1)} - B - \Omega_e  
\end{equation}
\begin{equation} 
\rho  = {\varepsilon \over c^2}  = {1 \over c^2}   
\Bigg\{ \Omega^{(0)} + \Omega^{(1)} + \sum_{f=u,d,s}\mu_f n_f   ~+ B + \Omega_e \Bigg\} \, , 
\end{equation}
with  $P = - \Omega$  being the pressure, $\rho$ the (mass) density and $\varepsilon$ the energy density.  

  The number densities for each particle species can be calculated using the 
thermodynamical relation: 
\begin{equation} 
       n_i = - \bigg({{\partial \Omega_i}\over{\partial \mu_i}}\bigg)_{TV}, 
     \qquad (i=u,d,s,e)
\end{equation}
and the  total baryon number density is 
\begin{equation}
       n   = {1\over{3}} (n_u + n_d + n_s)
\end{equation}
%...............................

For pedagogical purpose, let us consider the limiting case of massless non-interacting quarks 
($m_s = 0$, $\alpha_c = 0$). In this case SQM is composed by an equal number of $u$,$d$, $s$ 
quarks with no electrons ({\it i.e.} $n_e = 0$, $n_u = n_d = n_s,$) and the EOS takes 
the following simple form:  
\begin{equation}
        \varepsilon =  K n^{4/3} + B \, , ~~~~~~~~~~~
        P   = {1\over{3}} K n^{4/3} - B \, , ~~~~~~~~~~~
        K =   {9\over{4}} \pi^{2/3} \hbar c \, .
\label{bag_eos_00}
\end{equation}
Eliminating the total baryon number density, one has:
\begin{equation}
              P = {1\over{3}} (\varepsilon - 4B) \, . 
\label{bag_eos_0}
\end{equation}

To recapitulate, the properties of SQM, within this model, depend on the values of the three parameters 
$B$, $m_s$ and $\alpha_c$.  The EOS, in the form $P=P(\rho)$, is essentially determined by the value 
of the bag constant $B$.  The net fraction of leptons ($e^-$, $e^+$) which neutralize the electric 
charge of the quark component of SQM, will mainly depend on the values of $m_s$ and $\alpha_c$. 
Most frequently ({\it i.e.} for the most plausible values of the model parameters\cite{fj84}),  
the quark component has a positive electric charge, thus electrons will be present in SQM to neutralize it 
(as it was assumed by tacit agreement in the definition of SMQ given at the begining of this section)
\footnote{For small values of $m_s$ and large values of $\alpha_c$ the quark component of SQM 
has a negative charge\cite{fj84}, and thus positrons will be present to guarantee global 
charge neutrality.}. 

\subsection{The strange matter hypothesis}
According to the so called {\it strange matter hypothesis}\cite{bod71,ter79,witt84}   
(or Bodmer--Terazawa--Witten hypothesis), SQM is the true ground state of matter. 
In other words, the energy per baryon of SQM (at the baryon density where the pressure 
is equal to zero)  is supposed to be less than the lowest energy per baryon found in atomic 
nuclei, which is about 930.4 MeV for the most bound nuclei ($^{62}$Ni, $^{58}$Fe, $^{56}$Fe). 

If the strange matter hypothesis is true, then a nucleus with $A$ nucleons, could 
in principle lower its energy by converting to a {\it strangelet}.  
However, this process requires the simultaneous weak decay of about a number $A$ of $u$  
and $d$  quarks of the nucleus into strange quarks. The probability for such a process is 
proportional to $G_F^{2A}$, with $G_F$ being the Fermi constant.  Thus, for a large enough 
baryon number ($A > A_{min} \sim 5$), this probability is extremely low, and the mean life time 
for an atomic nucleus to decay to a strangelet is much higher than the age of the Universe.   
In addition, finite size effects (surface, coulomb and shell effects) place a lower limit 
($A_{min} \sim 10$--$10^3,$ depending on the values of the model parameters) on the 
baryon number of a stable strangelet even if in bulk SQM is stable\cite{fj84,madsen}.   
On the other hand, a step by step production ({\it i.e.} at different times) of {\it s} quarks 
will produce hyperons in the nucleus, that is to say, a system (hypernucleus) with a higher 
energy per baryon  with respect to the original nucleus. Thus according to the strange matter 
hypothesis, the ordinary state of matter, in which quarks are confined within hadrons is 
a metastable state with a mean life time much higher than the age of the Universe. 

The success of traditional nuclear physics, in explaining an astonishing amount of experimantal 
data, provides a clear indication that quarks in a nucleus are confined within protons and neutrons. 
Thus the energy per baryon for a droplet of {\it u}, {\it d} quark matter (the so called 
{\it nonstrange quark matter} in the bulk limit) must be higher than the energy per baryon of  
a nucleus with the same baryon number. 

These conditions in turn may be used to constrain the values of the parameters entering 
in models for  the equation of state of SQM and to find the region in the parameter 
space where the strange matter hypothesis is fulfilled and nonstrange quark matter is unstable. 
For example\cite{fj84,madsen}, in the case of the bag model EOS, for  non-interacting quarks 
($\alpha_c = 0$) one has $B \simeq $ 57--91 MeV/fm$^3$ for $m_s=0$, 
and $B \simeq $ 57--75 MeV/fm$^3$ for $m_s=150$ MeV.   
Our present understanding of the properties of ultra-dense hadronic matter, 
does not allow us to exclude or to accept {\it a priori} the validity of the strange matter hypothesis.

\section{Strange stars} 
%%%%%%%%%%%%%%%%%%%%%%%%5
One of the most important and captivating consequences of the strange matter hypothesis 
is the possible existence of {\it strange stars}, that is compact 
stars which are completely (or almost completely) made of SQM\footnote{When the values of the EOS parameters are such that the strange matter hypothesis is not fulfilled, the possible compact stars containing deconfined quark matter are the hybrid stars.}.   
These stars, as we will see in a moment, have bulk properties (mass  and radius)   
very similar to those of neutron stars (hadronic stars). 
Thus pulsars could be strange stars.   

The structural properties of non-rotating compact stars are obtained integrating numerically 
the Tolman--Oppenheimer-Volkoff (TOV) equations\cite{ov39,shap-teuk,gle_book}.     
The basic input to solve these equations is the stellar matter EOS. 
In the case of strange stars, one has to use one of the various models for SQM. 

The properties of the maximum mass configuration for strange star sequences obtained 
using a few  models for the SQM equation of state are reported in table 1. 
In this table, the EOS labeled as $B60_0$ refer to the bag model EOS (described in sect. 3) 
with $B=60$ MeV/fm$^3$ and $m_s = 0$;  
model $B60_{200}$ is for the same value of the bag constant but taking $m_s = 200$ MeV; 
model $B90_0$ refer to the case $B=90$ MeV/fm$^3$ and $m_s = 0$  
($\alpha_c = 0$ in all the three cases). The models denoted as  $SS1$ and $SS2$  refer to stellar sequences obtained with the equation of state for SQM  by Dey {\it et al.}\cite{dey98}  
%%%%%%%%%%%%%%%%%%%%%%%%%%%%%%%
\begin{table}       %Table~1
%%%%%%%%%%%%%%%%%%%%%%%%%%%%%%%
\tbl{Properties of the maximum mass configuration for strange stars obtained from different 
equations of state of SQM (see text for details). 
$M$ is the gravitational stellar (maximum) mass in unit of the solar mass $M_\odot$, $R$ 
is the corresponding radius, $\rho_c$ the central density, $n_c$ the central number density in unit of the saturation density ($n_0$ = 0.16 fm$^{-3}$) of nuclear matter, $P_c$ is the central pressure.}
{\begin{tabular}{@{}ccccccc@{}}
\hline\\
&&&&&\\[-15pt]
EOS & &$M/M_\odot$ & $R$   &  $\rho_c$    & $n_c/n_0$ & $P_c$   \\
       & &                    &(km)  & (g/cm$^3$) &                  & (dyne/cm$^2)$ \\[2pt] 
\hline\\
&&&&&&\\[-15pt]      
$B60_0$        && 1.964  &  10.71 & 2.06 $\times 10^{15}$ & 6.94 & 0.49 $\times 10^{36}$   \\[8pt]
$~B60_{200}$  && 1.751  &    9.83 & 2.44 $\times 10^{15}$ & 7.63 & 0.54 $\times 10^{36}$   \\[8pt]
$B90_0$        && 1.603  &    8.75 & 3.09 $\times 10^{15}$ & 9.41 & 0.73 $\times 10^{36}$   \\[8pt]
$SS1$            && 1.438  &    7.09 & 4.65 $\times 10^{15}$ &14.49 & 1.40 $\times 10^{36}$   \\[8pt]
$SS2$            && 1.324  &    6.53 & 5.60 $\times 10^{15}$ &16.34 & 1.64 $\times 10^{36}$   \\[8pt]
\hline     
\end{tabular} \label{SS_Mmax}} 
\end{table}               
%%%%%%%%%%%%%%%%%%%%%%%%%%%%%%

 In Fig.~\ref{sax_MR} we plot the mass-radius (MR) relation for strange stars (red curves) obtained 
using different model for SQM. For comparison, we plot in the same figure, the MR relations 
for hadronic stars (black curves): The curves labeled with $BBB1$, $BBB2$ (ref.\cite{BBB97}),   
$WFF$ (ref.\cite{wff88}) and $KS$ (ref.\cite{KS06}) are for {\it nucleon stars} ({\it i.e.} neutron stars having a $\beta$-stable nuclear matter core), the curve labeled $hyp$ refers to an 
{\it hyperon star} (ref.\cite{gm91,gle_book}).  
As we can see, there is a striking qualitative difference between the mass-radius relation of strange stars 
with respect to that of neutron stars. For strange stars with ``small'' ({\it i.e.} M not to close 
to $M_{max}$) mass, M is proportional to $R^3$. In contrast, neutron stars (hadronic stars) have radii 
that decrease with increasing mass. This difference in the MR relation is a consequence of the differences 
in the underlying interactions between the stellar constituents for the two types of compact stars. 
In fact, ``low'' mass strange stars are bound by the strong interaction, contrary to the case of neutron stars, 
which are bound by gravity
\footnote{As an idealized example, remember that pure neutron matter is not bound by nuclear forces 
(see {\it e.g.} ref.\cite{bl91}).}.      
This can be can be demonstrated looking at the different contributions (gravitational and internal 
binding energy) to the stellar total binding energy (see for example Fig.s 1, 2 and 3 in ref.
\cite{bomb01}). 

As it is well known, there is a minimum mass for a neutron star ($M_{min} \sim 0.1 M_\odot$). 
In the case of a strange star, there is essentially non minimum mass. As the stellar central density 
$\rho_c \rightarrow \rho_{surf}$ (surface density) a strange star (or better to say a lump of SQM 
for low value of the baryon number) is a self-bound system, until the baryon number becomes  
so low that finite size effects destabilize it.
%%%%%%%%%%%%%%% Fig. %%%%%%%%%%%%%
\begin{figure}[t]
\begin{center}
\psfig{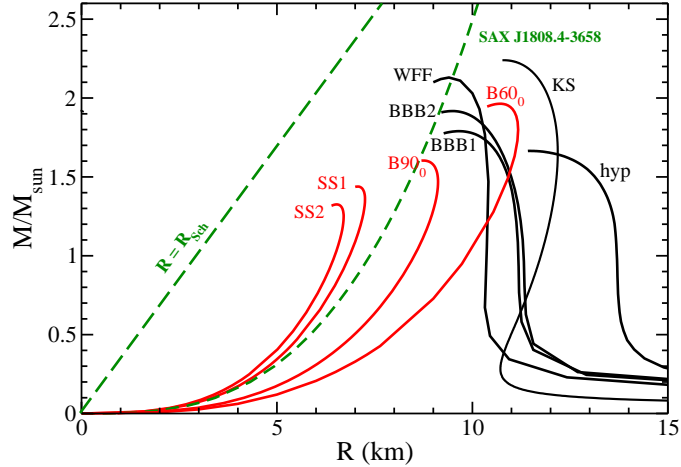}
\end{center}
\caption{The mass-radius relation for different theoretical models of compact stars.  
Black curves refer to hadronic stars and red curves to strange stars. The green dashed curve 
gives the upper limit, extracted from observational data by X.-D. Li {\it et al.}\cite{li99a},  
for the radius of the compact star in SAX~J1808.4-3658.  The green dashed straight line 
labeled $R_{Sch}$ gives the Schwarzschild radius as a function of the stellar mass.}    
\label{sax_MR}
\end{figure}
%%%%%%%%%%%%%%%%%%%%%%%%%%%%%%

\subsection{The surface: bare or crusted strange stars? } 
According to the standard view\cite{afo86,gw92,ste-mad05}, a strange star has a very sharp boundary.  
In fact, the density drops abruptly from 
$\rho_{surf} \sim 4$--$10 \times 10^{14}$ g/cm$^3$ to zero on a lenght scale typical of 
strong interactions, in other words the thickness of the stellar ``quark surface'' is of 
a few fermis  (1 fm = 10$^{-15}$m).  
This is of the same order of the thickness of the surface of an atomic nucleus. 
The density at the surface of a strange star, can be immediately calculated (in the limit 
$m_s \rightarrow 0$) using the simple EOS given in Eq.s~(\ref{bag_eos_00}) and (\ref{bag_eos_0}), 
and it is given by 
\begin{equation}
       \rho_{surf} = {{4 B}\over{c^2}} \, , ~~~~~~~~~~~~~~~~~~ n_{surf} =  \bigg ({{3 B}\over{K}}\bigg)^{3/4} \, . 
\label{surf_dens}
\end{equation} 
Strange stars with this sort of exposed quark matter surface are known as {\it bare} strange stars.    

Electrons are bound to the star by the electromagnetic force,  thus they can extend for several 
hundreds of fermis  above the ``quark surface''. This thin layer is usually reffered to as the 
{\it electrosphere}. As a consequence of this charge distribution a very strong electric field  
is established  at the stellar surface. This field has been estimated to be of about 
$10^{17}$ V cm$^{-1}$ and directed outward\cite{afo86}.   
Such a huge electric field is  expected to produce an intense emission of $e^+e^-$ pairs\cite{usov98} 
and a subsequent hard X-ray spectrum, at luminosities well above the Eddington limit
\footnote{The Eddington limit is the critical luminosity (due to matter accretion onto the star) 
above which the outward force due to the radiation pressure exceeds  
the inward gravitation force acting on  the infalling material. 
In the case of steady spherical accretion of hydrogen, Thomson scattering, Newtonian gravity,   
the  Eddington luminosity is given by\cite{shap-teuk} 
$L_{Edd} \simeq 1.3 \times 10^{38} (M/M_\odot)$~erg/s. 
A super-Eddington luminosity ($L > L_{Edd}$)  is allowed for a bare strange star,  because at 
its surface SQM is bound by strong interaction rather than gravity\cite{afo86,usov98}.}, 
as long as the stellar surface temperature is above\cite{usov01} $\sim 5 \times 10^8$~K.  
Thus a bare strange star should produce a striking characteristic signal, 
which differs both qualitatively and quantitatively from the thermal 
emission from a neutron star (hadronic star), and which  could provide an observational 
signature for their existence\cite{usov98,usov01,aks07,jai_etal04}.  
These special conditions ({\it i.e.} bare quark surface and high temperature) are realized 
in the case of ``young'' strange stars\cite{usov97,xu01,page-usov02}.  

Older and ``cold'' ($T <  5 \times 10^8$~K) strange stars will likely form a crust of 
``normal'' (nuclei and electrons) matter via mass accretion onto the  star 
from the interstellar medium or from a companion star, or from the matter left in the 
supernova explosion which could have formed the strange star. 
In fact, the strong  electric field at the stellar surface will produce a huge outward-directd force 
on any single positive ion (nucleus) of normal matter which is accreted onto the star. 
This force greatly overwhelms the force of gravity acting on the incoming positive ion. 
This accreted material will start to be accumulated on the top  of the electrosphere.  
Thus a strange star will form a crust of normal matter\cite{afo86}, which is suspended above 
the quark surface by the tiny ($\sim 10^2$--$10^3$~fm) electrostatic {\it gap}, and  
will completely obscure the ``quark surface''.  
This crust is similar in composition to the outer crust of a neutron star 
({\it i.e.} a Coulomb lattice of heavy neutron rich nuclei plus an electron gas)
In fact, when the density of matter at the base of this crust will reach the neutron drip density
($\rho_n^{drip} \simeq  4.3~10^{11}\times {\rm g/cm}^3$), the  neutrons could freely enter in the SQM 
stellar core, and there they will be dissolved (their constituent quarks will be deconfined) due to the assumed absolute stability of SQM.  
Thus, the density at the base of the crust can not exceeds the value of the neutron drip 
density. This condition sets an upper limit for the mass  of the stellar crust. 
For a strange star with a mass of 1.4~$M_\odot$,  the  mass of its crust is of the order of 
$M_{crust} \sim 10^{-5} M_\odot$ and its thickness of the order of a few hundreds of 
meters\cite{afo86,gw92}. 

Recently, the gap between the quark surface and the nuclear crust of strange stars has been 
investigated within a model\cite{ste-mad05} which accounts for the detailed balance between 
electrical and gravitational  forces and pressure, and in addition considers the effects 
of color superconductivity possibly occuring in the strange star core. 

These so called {\it crusted} strange stars will look very similar to neutron stars concerning 
their  emission properties, which are mainly determined by the stellar surface composition.  

Very recently, a new and alternative description of the surface region of a strange star has 
been proposed. According to the authors of ref.\cite{jrs06}, the crust of a strange star could 
consist of an heterogeneous phase composed of a Coulmob lattice of positively charged strangelets immersed in an sea of electrons. At the border with the uniforn SQM  stellar core, the crust 
will consit of bubbles filled with an electron gas embedded in SQM. In between these two 
extreme regions, one will find  various geometrical structures (rods, plates, etc) 
of one component (SQM/e$^-$ gas) into the other, similarly to the situation encounterd in 
the crust-core transition layer of a neutron star or in the mixed hadron-quark layer in a hybrid star.     
This altenative possibility for the crust structure, descend from imposing global 
(rather than local) electric charge  neutrality\cite{gle92},  and it could be realized when the 
finite-size energy cost (Coulmb, surface energies, etc) is less than the energy 
gain passing from the homogeneus to the hererogeneus phase. 
In the context of the bag model for SQM, it has been extimated\cite{alford06} that  
this  {\it strangelet crust}  could be formed when the quark matter surface tension 
is less that a critical value given by 
\begin{equation}
\sigma_{crit} \simeq 12\, \bigg(\frac{m_s}{150 \, {\rm MeV}} \bigg)^3 \,\frac{m_s}{\mu}~ {\rm MeV/fm}^2
\end{equation}
where $\mu$  is the quark chemical potential ($\mu \sim 300$ MeV in the surface region).   
For the {\it strangelet crust} of a strange star with a total mass of $1.4 M_\odot$ and total radius 
R = 10~km, the authors of ref. \cite{jrs06} have found $M_{crust} \simeq 6 \times 10^{-6} M_\odot$ 
and a radial extension $\Delta R \simeq 40$~m.  
In the case of a {\it strangelet crust}, there is a much reduced density jump at the 
stellar surface, and a different radiation spectrum is expectd  with respect to the case 
of a {\it bare} strange star.

\subsection{The core: role of color superconductivity } 
Until now, we have considered a very simple picture for the stellar core.  
The core is made of (locally) uniform SQM with a modest radial dependence of the density 
profile\cite{afo86}.  
However, the internal structure of a strange star could be much more complex. As we already 
mentioned in the introduction, a large number of color superconducting phases could be 
present in the range of densities spanned in the stellar core. Of particular astrophysical interst 
are cristalline phases of color superconducting SQM. The presence of these cristalline structures 
could constitute a basic prerequisite for modelling pulsar glitches in strange stars.

\section{Neutron star observations}

Many different kinds of astrophysical observations are currently 
used\cite{pag-red06,lat-pra06,klah06} to uncover the true nature of {\it neutron stars}. 
In the preceding pages, we have already mentioned  some of such observables. 
In addition, observations of the thermal radiation from  isolated neutron stars, combined with 
measurements (estimates) of their ages, allow for the determination of the stellar cooling  history. 
These data, when compared with theoretical cooling 
curves\cite{cool_1,cool_2,cool_3,cool_4,page-usov02},  
provide important  informations to get a realistic picture of the neutron star  composition. 

Binary stellar systems in which at least one component is a neutron star 
represent the most reliable way to measure the mass of the compact star.  
One of the most accurate mass determination is that of the neutron star 
associated to the pulsar PSR~1913 +16, which is a member of a tight 
(orbital period equal to 7 h 45 min) neutron star-neutron star system. 
The mass of PSR~1913 +16 is\cite{weis02}  $1.4408 \pm 0.0003 M_\odot$. 
Such impressive accuracy is made possible by measuring general relativistic 
effects, such as the orbital decay due to gravitational radiation, 
the advance of periastron, the Shapiro delay, {\it etc.}  
Neutron star masses in NS--NS  binary systems lie in 
the range 1.18 to 1.44 $M_\odot$  (see ref.\cite{kerk95,tc99}). 
Recently Nice {\it et al.} \cite{nial05}  have determined the mass 
of the neutron star associated to the millisecond pulsar PSR~J0751 +1808, 
which is a member of a binary system with a helium white dwarf secondary.  
Measuring general relativistic effects, the authors of ref. \cite{nial05} have obtained 
for the mass of PSR~J0751 +1808 the value $2.1 \pm 0.2 M_\odot$ at 68 \% confidence level 
($2.1^{+0.4}_{-0.5} M_\odot$ at 95 \% confidence level).  
Another important finding is the determination of a lower limit for the mass 
of the compact str in the pulsar I of the globular cluster Terzan 5 (Ter~5~I). 
At 95\% confidence, the mass of Ter~5~I exceeds \cite{raal05}  1.68 $M_\odot$.  
These {\it large} values of the measured  masses of PSR~J0751 +1808 and Ter5~I are 
very important, since they push up (with respect to the previous measured mass 
sample\cite{kerk95,tc99}) the  lower limit of the value of the Oppenheimer-Volkoff maximum 
mass $M_{max}$ of neutron stars. As it is well known, $M_{max}$ is directlty related to 
the overall {\it stiffness} of the EOS.  

Another quantity related to the  EOS is the maximum rate of rotation ($\Omega_{max}$) 
sustainible by a  compact star.  Equilibrium sequences of rapidly rotating compact stars have been constructed numerically in general relativity by several groups\cite{bonazz93,cst94,dtb98,btd00}.   
The numerical results for $\Omega_{max}$  obtained for a broad set of realistic EOS 
can reproduced with a very good accuracy using simple empirical formulae 
which relate $\Omega_{max}$ to the mass ($M_{max}$) and radius $R_0$  of the 
non-rotating maximum mass configuration\cite{sal94}. 

In the following part of this section, we discuss a few observational constraints for 
the mass-radius  relation of neutron stars.

\subsection{SAX~J1808.4-3658}
The transient X-ray source and Low Mass X-ray Binary (LMXB)  SAX~J1808.4-3658 
was discovered in September 1996 by the Dutch-Italian BeppoSAX satellite\cite{zand98}   
Type-I X-ray bursts have been detected from this source in September 1996, April 1998, 
and October 2002. 
Very recently, a weak precursor event to the burst of October 19,  2002, has been 
identified\cite{sudip07} in the observational data taken by the Rossi X-ray Timing Explorer satellite.   
Coherent X-ray pulsation with a period of $P = 2.49$ ms  was  discovered in 1998 by Wijnands and 
van der Klis \cite{wij-klis_98}.    
Lately, {\it burst oscillations} ({\it i.e.} millisecond-period brightness oscillations during X-ray bursts)  
have been detected\cite{chakr03} for this source, at the frequency of $\sim 401$~Hz, 
{\it i.e.} at the same frequency of the coherent X-ray pulsation. This has confirmed the 
expectation that the burst oscillations frequency is equal to the pulsar frequency, and thus to 
the rotational spin frequency of the associated compact star.  

SAX~J1808.4-3658 is the first and the best studied member of the class of 
{\it accretion-driven millisecond pulsars} (see \cite{stroh01,wijn06} and references therein quoted).  
These sources are believed to be the progenitors of millisecond radio pulsars by the way 
of spin up by mass transfer from the companion star in a LMXB\cite{Bhat-vdH_91}.

Using the measured X-ray fluxes during the high- and low-states of the source 
in the time of the April 1998 outburst, and in addition the restrictions from modelling the 
observed X-ray pulsation at  $P = 2.49$ ms,  X.-D. Li {\it et al.}\cite{li99a}  have derived 
an upper limit for the radius of the compact star in SAX~J1808.4-3658, which is given by the 
dashed curve in Fig.~\ref{sax_MR}. The dashed line, labeled $R = R_{Sch}$, 
in this figure represents  the Schwarzschild radius, which is the lower limit for the stellar radius. 
In fact, being a source of type-I X-ray burst and according to the current interpretetion of this phenomenon (thought to be produced by thermonuclear explosion on the surface of a 
{\it neutron star}), the compact object in SAX~J1808.4-3658 can not be a black hole.  
Thus the allowed range for the mass and radius of SAX~J1808.4-3658 is the region confined 
by these two dashed lines in Fig.~\ref{sax_MR}. 
In the same figure,  we report the theoretical mass-radius relations for different sequences 
for pure hadronic compact star (black curves) and strange stars. These models have been 
illustrated in the previous section. The results reported in Fig.~\ref{sax_MR}, clearly indicate 
that a strange star model is more compatible with the semiempirical mass-radius relation\cite{li99a} 
for  SAX~J1808.4-3658, than an hadronic star  model.  

Recently Leahy {\it et al.}\cite{leahy07}, from the analysis of the light curves of 
SAX~J1808.4-3658 during its 1998 and 2002 outburst, have obtained limits on the mass and 
radius for the compact star in this source. At the 99.7\% confidence level they obtain 
$6.9~{\rm km} \leq R \leq 11.9$ km, and $0.75  \leq M/M_{\odot} \leq 1.56$. 
These limits are compatible with strange stars models as well as with hybrid or pure hadronic 
stars (see the theoreticla MR curves in Fig.s \ref{sax_MR} and \ref{xte_MR}). 
For example,  in the case of a non-rotating {\it nucleonic stars}  within the  BBB1 
equation of state for a star with $M = 1.0 M_\odot$,  one has $R = 11.2$ km. 
Rotation with a period of 2.5 ms, will not change appreciabily the value of the stellar radius, 
since this rotation rate is still "far" from the mass shed limit for this EOS\cite{dtb98}. 
 
%%%%%%%%%%%%%%% Fig. %%%%%%%%%%%%%
\begin{figure}[t]
\begin{center}
\psfig{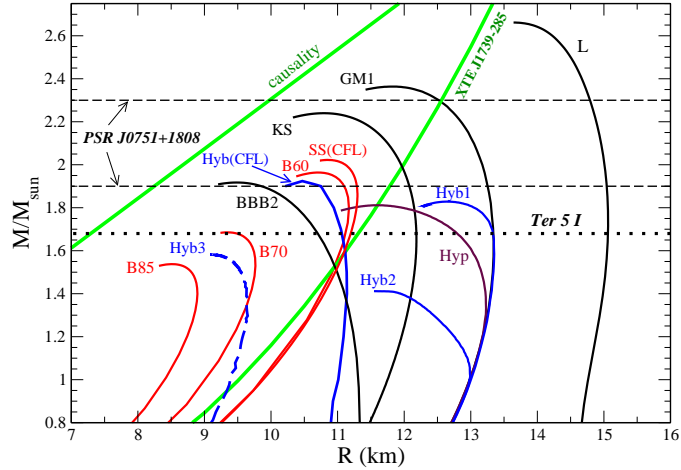}
\end{center}
\caption{Mass-radius plane with limits from the rotational  frequency at 1122 for 
XTE~J1739-285 and causality (green lines).  
Various theoretical MR relations are reported (see text for details). 
The dashed horizontal line denotes the measured mass (at 68 \% confidence level)  of the 
compact star in PSR~J0751 +1808.  The dotted line gives the lower limit (at 95\% confidence 
level) for the mass of the compact star in the pulsar I of the globular cluster Terzan 5.}  
\label{xte_MR}
\end{figure}
%%%%%%%%%%%%%%%%%%%%%%%%%%%%%%

\subsection{XTE~J1739-285}  
In a recent paper Kaaret {\it et al.}\cite{kaar07}  have reported the discovery of 
burst oscillations at 1122 Hz in the X-ray transient XTE~J1739-285.    
If the  burst oscillation frequency in this source is coincident with the stellar spin rate, 
as in the case of SAX~J1808.4-3658 (ref.\cite{chakr03}), thus  XTE~J1739-285  contains 
the most rapidly rotating compact star discovered up to now,  and the first with a 
submillisecond spin period ($P = 0.891$ ms).  
This discovery gives a  model indipendent observational contraint to EOS dense stellar matter, 
and it has prompted a number of studies\cite{lavag06,bejg07,drago07}  in this direction. 

In one of these studies, Lavagetto {\it et al.}\cite{lavag06} have derived the following upper limit 
\begin{equation}
                 R < 9.52 \big(M/M_\odot \big)^{1/3} ~{\rm km}
\end{equation} 
for the radius of the compact star in XTE~J1739-285, using its inferred spin period and 
a simple empirical formula\cite{latt06}  which appriximately gives the minimum rotation period 
$P_{min} = 2 \pi/\Omega_{max}$ for a star of massa $M$ and non-rotating radius $R$. 
This  upper limit is depicted in  Fig.~\ref{xte_MR} by the green curve labeled XTE~J1739-285. 
For comparison are reported in the same figure, the theorethical MR relations 
for nucleon stars (black curves), hyperon stars (brown curve), hybrid stars (ble curves) 
and strange stars (red curves). Some of these models have been already mentioned in the 
previous sections. Curve (L) is relative to neutrons stars with a pure neutron matter core\cite{ab77}. This stellar seuqence has been included as an extreme example of a very stiff EOS.  
The nucleon star sequence labeled (GM1) has been obtained within a relativistic 
quantum field theory approch in the mean field approximation, for one of the parameter 
set (GM1)  given by  Glendenning and Moszkowski\cite{gm91}.  
The curve (Hyp) referes to hyperon stars  calculated with the previous GM1 EOS when 
hyperons are include. Curves (Hyb1) and (Hyb2) represent hybrid star sequences 
calculated using the same GM1 EOS for hyperonic matter to describe the hadonic phase, 
and the bag model EOS  to describe the quark phase.  We took  $B = 208.24$~MeV/fm$^3$ (Hyb1), 
$B = 136.63$~MeV/fm$^3$ (Hyb2) and $m_u = m_d =0$, $m_s = 150$~MeV. 
The curve (Hyb3) refers to hybrid stars constructed using a different 
parametrization for  the EOS by \cite{gm91} (GM3) for the hadronic phase 
and  using $B = 80$~MeV/fm$^3$. 
The MR curve labeled (Hyb(CFL)) refers to hybrid stars \cite{alfal05} whose core is a mixed 
phase of nuclear matter and  color-superconducting CFL quark matter.      
Finally, we consider CFL strange stars\cite{lug03} (SS(CFL)) within the bag model EOS  
with  $B = 70$~MeV/fm$^3$,  $m_u = m_d = 0$ and $m_s = 150$ MeV and  
quark pairing gap $\Delta = 100$ MeV.  

From the results in Fig.~\ref{xte_MR}, we see that the constraint derived from the burst 
oscillations in XTE~J1739-285 does not allow to discriminate among possible different types 
of compcat stars.  Anyhow, no one of the present astrophysical observations can 
prove or confute the existence of strange stars (or hybrid stars), {\it i.e.} the presence 
of SQM compact stars\cite{pag-red06,lat-pra06,klah06,alf_etal06}.

\section{Metastability of hadronic stars and their delayed conversion to quark stars} 

One of the most recent developments in studying the astrophysical implications of   
SQM in compact stars is the realization that pure hadronic compact stars above a   
threshold value of their mass are metastable\cite{berez03,3i,drago04}.  
The metastabity of hadronic stars originates from the finite size effects in the formation 
process of the first SQM drop in the hadronic enviroment.  

In cold ($T$ = 0) bulk matter the quark-hadron mixed phase begins at the 
{\it static transition  point} defined according to the Gibbs' criterion 
for phase equilibrium 
\begin{equation}
\mu_H = \mu_Q \equiv \mu_0 \, , ~~~~~~~~~~~~
P_H(\mu_0) = P_Q(\mu_0) \equiv P_0  \,     
\label{eq:eq1}
\end{equation}
where $ \mu_H = (\varepsilon_H + P_H)/n_{b,H}$  and $\mu_Q = (\varepsilon_Q + P_Q)/n_{b,Q}$   
are the chemical potentials for the hadron and quark phase respectively, 
$\varepsilon_H$ ($\varepsilon_Q$),  $P_H$ ($P_Q$)  and $n_{b,H}$  ($ n_{b,Q}$)
denote respectively the total ({\it i.e.,}  including leptonic contributions) energy 
density,  the total pressure and baryon number density  for the hadron (quark)  
phase.      

Consider now  the more realistic situation in which one takes into account 
the energy cost due to finite size effects  in creating a drop of deconfined QM  
in the hadronic environment.  As a consequence of these effects,  the formation 
of a critical-size drop of QM is not immediate and it is necessary to have 
an overpressure  $\Delta P =  P - P_0$  with respect to the static  transition point.  
Thus,  above  $P_0$, hadronic matter is  in a metastable state,  and the formation 
of a real drop of QM occurs via a quantum nucleation mechanism.  
A sub-critical (virtual) droplet of deconfined QM moves back and 
forth in the potential energy well separating the two matter phases on a time scale 
$\nu_0^{-1} \sim 10^{-23}$ seconds, which is set by the strong interactions. 
This time scale is many orders of magnitude shorter than the typical time scale 
for the weak interactions,  therefore 
quark flavor must be conserved during the deconfinement transition. 
We will refer to this form of deconfined matter,  in which the 
flavor content is equal to that of the $\beta$-stable hadronic system 
at the same pressure,  as the  Q*-phase. 
Soon afterwards a critical size drop of QM  is formed the weak interactions 
will have enough time to act, changing the quark flavor fraction of the deconfined 
droplet to lower its energy, and a droplet of $\beta$-stable SQM is formed 
(hereafter the Q-phase).

%%%%%%%%%%%%%%% Fig. %%%%%%%%%%%%%
\begin{figure}[t]
\begin{center}
\psfig{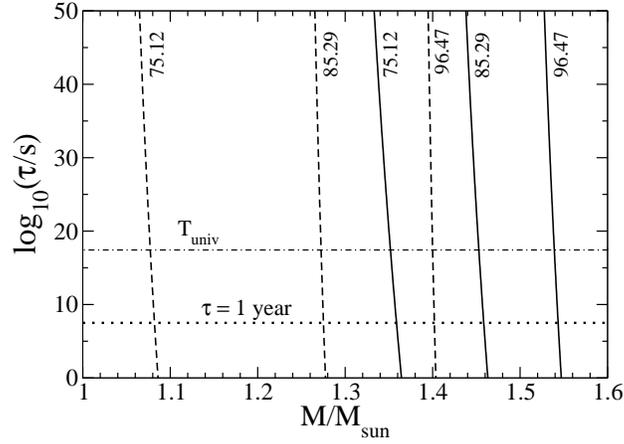}
\end{center}
\caption{Nucleation time as a function of the  gravitational mass of the 
hadronic star. Solid lines correspond to a value of $\sigma=30$ MeV/fm$^2$ whereas 
dashed ones are for $\sigma=10$ MeV/fm$^2$.     
The nucleation time corresponding  to one year  is shown by the dotted horizontal line. 
The different values of the bag constant (in units of MeV/fm$^3$) are plotted next 
to each curve. The hadronic phase is described with the GM1 model for 
the equation of state.}
\label{fig:tau} 
\end{figure}
%%%%%%%%%%%%%%%%%%%%%%%%%%%%%%

In the scenario proposed by the authors of ref.\cite{berez03}, a pure  HS  whose central 
pressure  is increasing due to spin-down or due to mass accretion, {\it e.g.,} from the 
material left by the supernova explosion,  or from a companion star.    
As the central pressure  exceeds  the threshold value $P_0^*$ at the static 
transition point,  a virtual drop of quark matter in the Q*-phase can be formed 
in the central region of the star.   As soon as a real drop of Q*-matter is formed, 
it will grow very rapidly and the original HS will be converted to strange star or to 
an hybrid star, depending on whether or no the strange matter hypothesis is fulfilled.  

To calculate the nucleation time $\tau$, {\it i.e.,} the time needed to form the first  
critical droplet of deconfined QM  in the hadronic medium,   
one can use  the relativistic quantum nucleation theory \cite{is97} 
(for more details,  see \cite{berez03,3i}).     
The nucleation time $\tau$,  can be calculated for different values of the stellar 
central pressure $P_c$.  In Fig.\ \ref{fig:tau},  we plot $\tau$ as a function of the 
gravitational mass $M_{HS}$ of the HS corresponding to the given value of the 
central pressure, as implied by the solution of the TOV equations for the pure HS sequences.  
Each curve refers to a different value of  the bag constant and  surface tension $\sigma$.    
As we can see,  from the results in Fig.\  \ref{fig:tau}, a metastable  hadronic star can 
have a mean-life time many orders of magnitude larger than the age of the universe  
$T_{univ} = (4.32 \pm 0.06) \times 10^{17}$~s.  As the star accretes a small amount 
of mass  (of the order of a few per cent of the mass of the sun),  
the consequential increase of the central pressure leads  to a huge 
reduction of the nucleation time and, as a result, to a dramatic reduction 
of the HS {\it mean-life time}.  
%%%%%%%%%%%%%%% Fig. %%%%%%%%%%%%%
\begin{figure}[t]
\begin{center}
\psfig{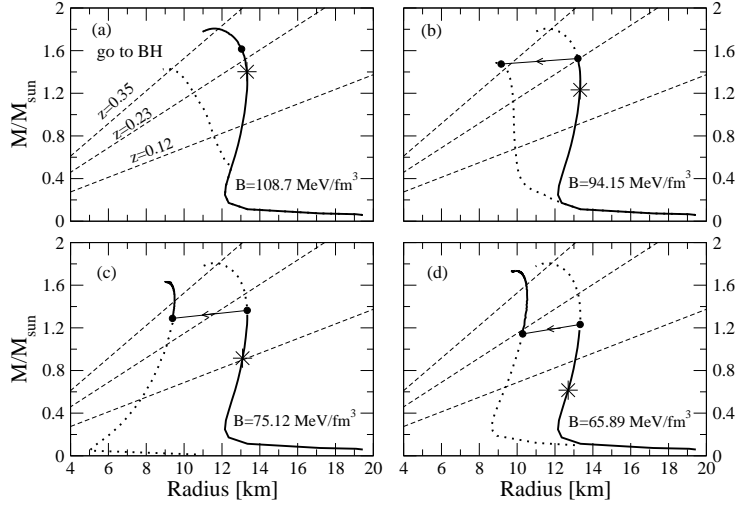}
\end{center}
\caption{Mass-radius relation for  pure HS described within the GM1 model and 
for  HyS or SS configurations for several values of the bag constant.  
The configuration marked with an asterisk represents the HS for which the 
central pressure is equal to $P_0^*$ and $\tau = \infty$.   
The conversion process of the HS, with a mass equal to $M_{cr}$,  into
a final QS is denoted by the full circles connected by an arrow. 
In all the panels $\sigma$= 30 MeV/fm$^2$.  
The dashed lines show the gravitational red shift deduced for the X-ray compact 
sources EXO~0748-676  ($z = 0.35$) and 1E~1207.4-5209  ($z =$ 0.12 -- 0.23).
See ref.\cite{3i} for more details and references to the red shift data.}   
\label{fig:MR}
\end{figure}
%%%%%%%%%%%%%%%%%%%%%%%%%%%%%%

To summarize,  pure hadronic stars having  a central 
pressure larger than the static transition pressure  for the formation 
of the Q*-phase are metastable to the ``decay'' (conversion) to a more compact 
stellar configuration in which deconfined QM  is present 
(HyS or SS). These metastable HS  have a {\it mean-life time}  which is 
related to the nucleation time to form the first critical-size drop of deconfined 
matter in their interior
%%%%%%%%%%%%%%%
{\footnote{~The actual  {\it mean-life time} of the HS will depend on the 
mass accretion or on the spin-down rate which modifies the nucleation time via  
an explicit time dependence of the stellar central pressure.}}. 
%%%%%%%%%%%%%%%%%%%%%%%%%%%%%%%%%%%%%%%%%%%%%%%%%  
The {\it critical  mass} $M_{cr}$ of the metastable HS is defined\cite{berez03,3i}  as the 
value of the  gravitational mass for which the nucleation time is equal to one year: 
$M_{cr}\equiv M_{HS}(\tau$=1\,yr).    
Pure hadronic stars with $M_H > M_{cr}$ are very unlikely to be observed.  
$M_{cr}$  plays the role of an {\it effective maximum mass} 
for the hadronic branch of compact stars.      
While the Oppenheimer--Volkov maximum mass \cite{ov39} $M_{HS,max}$   
is determined by the overall stiffness of the EOS for hadronic matter, 
the value of $M_{cr}$  will depend in addition on the bulk properties of the EOS 
for quark matter and on the properties at the interface between  
the confined and deconfined phases of matter ({\it e.g.,} the surface tension $\sigma$). 

In Fig.\ \ref{fig:MR},  we show the mass-radius (MR) curve for pure HSs  within 
the GM1 model for the EOS of the hadronic phase, and that for hybrid or strange 
stars for different values of the bag constant $B$.     
The configuration marked with an asterisk on the hadronic MR curves represents 
the HS  for which the central pressure is equal to $P_0^*$ and $\tau = \infty$.  
The full circle on the HS sequence represents the critical mass configuration, 
in the case  $\sigma = 30$ MeV/fm$^2$.    
The  full circle on the HyS (SS) mass-radius curve represents the hybrid (strange) star  
which is formed from the conversion of the hadronic star with $M_{HS}$=$M_{cr}$. 
We assume  that during the stellar conversion process the total number of baryons 
in the star ({\it i.e.} the stellar baryonic mass)  is conserved.  Thus the total energy 
liberated in the stellar conversion is given \cite{bd00} by the difference between 
the gravitational mass of the initial hadronic star  ($M_{in} \equiv M_{cr}$)  
and that of the final hybrid or strange stellar configuration  
$M_{fin} \equiv M_{QS}(M^b_{cr})$ with the same baryonic mass:  
         $ E_{conv} = (M_{in} - M_{fin}) c^2$.  

The stellar conversion process starts to populate the new branch of quark stars 
(see Fig.\ \ref{fig:MR}).     
Long term accretion on the quark star can next produce stars with 
masses up to the maximum mass $M_{QS,max}$ for the QS sequence.

The possibility to have metastable hadronic stars, together with the feasible 
existence of two distinct families of compact stars, demands an extension of the 
concept of maximum mass of a ``neutron star'' with respect to 
the {\it classical} one introduced by Oppenheimer and Volkoff\cite{ov39}.     
Since metastable HS with a ``short'' {\it mean-life time} are very unlikely to be observed,  
the extended concept of maximum mass must be introduced in view of the comparison 
with  the values of the mass of compact stars deduced from direct astrophysical 
observation.  
Having in mind this operational definition, the authors of ref.\cite{3i} define as 
{\it limiting mass} of a compact star, and denote it as $M_{lim}$, 
the physical quantity defined in the following way: 

\noindent 
({\it a})  if the nucleation time $\tau(M_{HS,max})$  associated to the maximum mass 
configuration for the hadronic star sequence is of the same order or much larger  than  
the age of the universe $T_{univ}$,   then 
\begin{equation}
         M_{lim}  =  M_{HS,max} \, ,  
\end{equation}
in other words, the limiting mass in this case coincides with the Oppenheimer--Volkoff 
maximum mass for the hadronic star sequence. 

\noindent 
({\it b}) If the critical mass $M_{cr}$  is smaller than $M_{HS,max}$  
({\it i.e.}  $\tau(M_{HS,max}) < 1$~yr), 
thus the limiting mass for compact stars is equal to the largest value between the 
critical mass for the HS and the maximum mass for the quark star (HyS or SS) sequence 
\begin{equation}
         M_{lim} =   max \big[M_{cr} \, ,  M_{QS,max} \big] \, .
\end{equation}

\noindent 
({\it c}) Finally, one must consider an ``intermediate'' situation for which      
                          $1 {\rm yr} <  \tau(M_{HS,max})  <  T_{univ}$.  
As the reader can easely  realize, now 
\begin{equation}
         M_{lim} =   max \big[M_{HS,max} \, ,  M_{QS,max} \big] \, , 
\end{equation}
depending on the details of the EOS which could give $M_{HS,max} > M_{QS,max}$ 
or vice versa.  

The delayed stellar conversion process, described so far, represents 
the second ``explosion''  -- the {\it Quark Deconfinement Nova} (QDN) -- 
in the two-step scenario proposed in  ref. \cite{berez03,3i}      
to explain a possible delayed SN-GRB connection.    
  
%%%%%%%%%%%%%%%%%%%%%%%%%%%%%%%%%%%%%%%%%%%%%%%%%%

\vfill
%%%%%%%%%%%%%%%%%%%%%%%%%%%%%%
\end{document}